\documentclass[]{spie}  

\usepackage{amssymb}
\usepackage{amsmath}
\usepackage{graphicx}
\usepackage{booktabs}
\usepackage{siunitx}
\usepackage{xcolor}
\usepackage[colorlinks=true, allcolors=blue, breaklinks=true]{hyperref}  

\newcommand{\degree}{$^{\circ}$}

\title{FOSSIL's preliminary thermal architecture}

\author[a]{Valentin Sauvage}
\author[a]{Clémence de Jabrun}
\author[a]{Anaïs Besnard}
\author[a]{Bruno Borgo}
\author[b]{Ivan Charles}
\author[b]{Jean-Marc Duval}
\author[a]{Bruno Maffei}
\author[b]{Sylvain Martin}
\author[a]{Nabila Aghanim}
\affil[a]{Univ. Paris Saclay, CNRS, Institut d'Astrophysique Spatiale, Building 121, Orsay, France}
\affil[b]{Univ. Grenoble Alpes, CEA, IRIG, DSBT, 38000, Grenble, France}

\authorinfo{Send correspondence to Valentin Sauvage: E-mail: valentin.sauvage@universite-paris-saclay.fr}

\begin{document}
\maketitle

\begin{abstract}
FOSSIL (FTS fOr CMB Spectral diStortIon expLoration) is a proposed ESA M8 mission tailored to measure the spectral distortions monopole of the Cosmic Microwave Background (CMB) with a sensitivity three orders of magnitude beyond the COBE/FIRAS legacy measurement. Achieving this sensitivity demands an extraordinarily challenging cryogenic architecture: the scientific instrument must be maintained at 4.5~K, while the detector focal plane assembly operates at 50~mK. This paper presents an overview of the preliminary thermal architecture of the FOSSIL payload, from the spacecraft service module at 293~K down to the sub-kelvin detector stage. The design draws on heritage from the Planck and ARIEL missions and relies on a staged passive cooling chain comprising a multi-layer insulation blanket, three V-groove radiators (operating at approximately 130~K, 90~K, and 50~K), and a 25~K actively cooled shield fed by an ESA-provided 4~K mechanical cryocooler. Sub-kelvin temperatures are achieved via a multi-stage adiabatic demagnetisation refrigerator (ADR) developed for NewAthena/X-IFU, providing continuous cooling at 1.8~K and 350~mK, and 50~mK with an 80\% duty cycle. The Focal Plane Assembly (FPA), housing four Kinetic Inductance Detector (KID) arrays at 50~mK, is thermally isolated from the 4.5~K bench via a carbon-fibre reinforced polymer (CFRP) hexapod with staged heat interception. We outline the staged cooling concept and show that the architecture closes with positive thermal margins at every stage; the detailed steady-state thermal budget will be presented in a forthcoming dedicated paper.
\end{abstract}

\keywords{CMB spectral distortions, satellite thermal architecture, adiabatic demagnetisation refrigerator}

\section{INTRODUCTION}
\label{sec:intro}

The measurement of the absolute frequency spectrum of the Cosmic Microwave Background (CMB) represents one of the most powerful probes of early-Universe physics.\cite{Fixsen1996, Voyage2050SDWP} Since the landmark COBE/FIRAS measurement\cite{Fixsen1996, Fixsen2009} and their recent re-analysis,\cite{fabbian2025newconstraintydistortionfiras} the CMB spectrum has been known to be an exquisitely precise blackbody. A broad class of physical processes, from the dissipation of primordial density perturbations to the integrated thermal history of large-scale structure formation, inevitably imprint small deviations from a pure Planck spectrum, known as spectral distortions.\cite{Sunyaev1970a, Sunyaev1970b} These remain undetected, lying three orders of magnitude below the FIRAS sensitivity floor, and constitute a unique guaranteed science target for the next generation of CMB spectrometers.

FOSSIL (FTS fOr CMB Spectral diStortIon expLoration) is a mission concept submitted in response to the ESA M8 call for proposals,\cite{fossil_general_paper} and draws on the heritage of PIXIE\cite{pixie2011,Kogut2020PIXIE} and earlier concepts. It aims to map the full sky from 50~GHz to 2~THz using a Martin-Puplett Fourier Transform Spectrometer (FTS), providing 130 spectral channels at a resolution of 15~GHz, with primary science goals of detecting the $\mu$-type distortion monopole at $\sigma(\mu) \simeq 1.5 \times 10^{-8}$ and the $y$-type monopole at $\sigma(y) \simeq 5 \times 10^{-9}$.\cite{coulon_paper_sd, fossil_general_paper}

Measuring these faint distortions places some of the most stringent cryogenic requirements of any astrophysical discipline, requiring three conditions to be met simultaneously, which no existing or previously flown mission has achieved together: ultra-sensitive detectors operating at 50~mK; an instrument at $\sim$4.5~K to suppress its own thermal emission to a level manageable by calibration; and an absolute calibration reference stabilised to within a few $\mu$K at temperatures bracketing the CMB monopole at $T_0 \simeq 2.7255$~K. This paper presents an overview of the preliminary thermal architecture developed to meet these requirements, from the service module interface at ambient temperature down to the sub-kelvin detector stage, building on the staged passive cooling demonstrated on Planck\cite{Planck_thermal2011} and ARIEL\cite{ARIEL_thermal2022}, the actively cooled shield architecture of SPICA\cite{Saijo2021_SPICA_CryogenicCooling} and the NG-CryoIRTel CDF study,\cite{NG_CryoIRTel2014} and the multi-stage ADR integrated within the cryo chain demonstration for the previous ATHENA version\cite{Prouve2020}. At the present early stage of the project, the architecture described here is a viable baseline concept rather than a consolidated design, intended to demonstrate overall feasibility ahead of the detailed trade-offs and optimisation of Phase~A.

\section{THERMAL REQUIREMENTS AND DESIGN PHILOSOPHY}
\label{sec:requirements}

The thermal requirements of FOSSIL flow directly from its science objectives. The fundamental measurement principle is a differential comparison between sky emission and an actively cooled blackbody internal reference (BBIR) using the FTS. To suppress instrumental self-emission, the optimal instrument temperature is as close as possible to the CMB monopole temperature; a practical upper limit of 4.5~K is adopted for the instrument stage, enabled by a 4~K mechanical cryocooler currently under development at ESA.

The detector units must operate at 50~mK for two reasons. First, the KID materials must operate well below their superconducting critical temperature $T_c$: for the LF band (50--300~GHz), the absorbing resonators require $T_c \sim 0.7$~K, achieved through the proximity effect in a tri-layer Al-Ti-Au thin film;\cite{Catalano2015,Catalano2020} operating at $T \lesssim T_c/10$ ensures a stable superconducting regime. Second, achieving the photon-noise-limited regime requires that the detector NEP$_\mathrm{det} \lesssim 3 \times 10^{-17}$~W~Hz$^{-1/2}$ remain at least one order of magnitude below the mean photon noise. State-of-the-art KIDs have demonstrated NEP$_\mathrm{det} \sim 3 \times 10^{-20}$~W~Hz$^{-1/2}$,\cite{Baselmans2022} comfortably exceeding this requirement.

\begin{figure}[ht]
\centering
\includegraphics[width=0.85\linewidth]{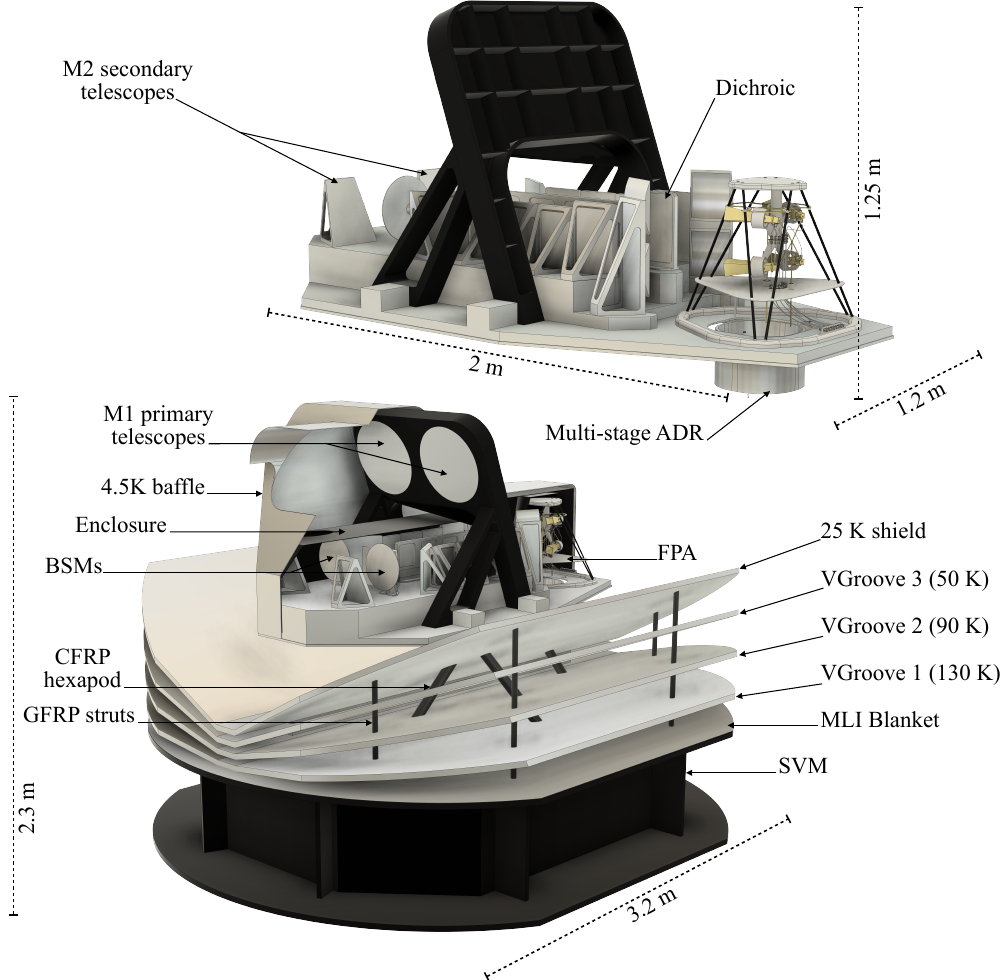}
\caption{Overall view of FOSSIL on the spacecraft, benchmarked and upscaled from ARIEL's platform (bottom), and view of the FOSSIL instrument with its main elements identified (top). The staged cryogenic chain runs from the service module (SVM) through the MLI blanket and three V-grooves to the 25~K shield, the 4.5~K instrument bench, and the sub-kelvin focal plane assembly cooled by the multi-stage ADR.}
\label{fig:overview}
\end{figure}

The thermal architecture follows a layered approach, with each cooling stage progressively reducing the load intercepted by the following one (Table~\ref{tab:stages}), mirroring the approach employed on Planck.\cite{Planck_thermal2011} An overall view of the payload and its cryogenic stages is shown in Figure~\ref{fig:overview}. The architecture maximises the benefit of the L2 orbit, which provides an exceptionally stable thermal environment with a tightly constrained solar aspect angle of approximately $\pm$10\degree. The heritage of Planck demonstrates the thermal stability achievable at L2, where the HFI bolometer plate was actively stabilised below $20~\mathrm{nK}~\mathrm{Hz}^{-1/2}$ in the science band and the 4~K optical stage met a $10~\mu\mathrm{K}~\mathrm{Hz}^{-1/2}$ requirement.\cite{Planck_thermal2011, Pajot2010}

\begin{table}[ht]
\centering
\caption{Summary of the FOSSIL thermal stages and associated cooling means.}
\label{tab:stages}
\begin{tabular}{lll}
\toprule
\textbf{Stage} & \textbf{Temperature} & \textbf{Cooling Mean} \\
\midrule
SVM panel        & 293~K          & Passive (solar panels and radiators) \\
V-Groove 1       & $\sim$130~K    & Passive radiation \\
V-Groove 2       & $\sim$90~K     & Passive radiation \\
V-Groove 3       & $\sim$50~K     & Passive radiation \\
25~K shield      & $\sim$25~K     & Active (4~K cooler) \\
Instrument bench & 4.5~K          & Active (4~K cooler) \\
BBIR             & 2.5--2.9~K     & Multi-ADR T2 \\
ADR T2 stage     & 1.8~K (cont.)  & Multi-ADR \\
ADR T1 stage     & 350~mK (cont.) & Multi-ADR \\
FPA / Detectors  & 50~mK          & Multi-ADR T0 stage \\
\bottomrule
\end{tabular}
\end{table}

\section{PASSIVE CHAIN: MLI, V-GROOVES AND 25~K SHIELD}
\label{sec:passive}

The passive cooling chain draws directly on the Planck heritage,\cite{Planck_thermal2011} whose three V-groove radiators and MLI architecture demonstrated purely passive cooling from the SVM temperature down to $\sim$50~K at L2. A 20-layer MLI blanket covering the SVM top surface forms the first thermal barrier between the service module at $\approx$293~K and the payload. Three V-groove radiators, inclined at increasing angles with respect to the spin axis, then progressively intercept and re-radiate heat toward deep space, reaching equilibrium temperatures of approximately 130~K, 90~K and 50~K. The panels are modelled as aluminium honeycomb sandwich structures following the ARIEL heritage,\cite{ARIEL_thermal2022} with a surface treatment strategy (VDA hot faces, low-emittance cold faces, high-emissivity space-facing surfaces) validated on Planck. The structural support struts between the SVM and each V-groove stage are made of GFRP G-10, with heat intercepted at each stage via collars and braided straps.

On top of the passive stack sits the 25~K shield, actively cooled by the ESA-provided 4~K cooler chain. It provides a further interception stage before the 4.5~K instrument and thermally pre-conditions the Low Noise Amplifiers (LNAs) of the KID readout, which operate between 4 and 30~K. The structural support between the SVM, the 25~K shield and the 4.5~K bench is provided by a hexapod of six CFRP struts, sharing the ARIEL payload support heritage.

\section{THE 4.5~K INSTRUMENT STAGE}
\label{sec:4K}

The FOSSIL instrument is mounted on a 4.5~K aluminium bench surrounded by a light-tight Aluminium baffle. The bench houses all optical and quasi-optical elements: the two off-axis dual-mirror telescopes (primary 420~mm, secondary 200~mm), the Martin-Puplett FTS with its scanning mirror mechanism, the BBIR, the two beam-switching mechanisms (BSMs), the dichroic beam splitters, and the FPA. Cooling the full enclosure to 4.5~K minimises the instrumental self-emission, reducing the differential signal between the instrument and the sky to a level manageable by the BBIR-based calibration scheme (Section~\ref{sec:bbir}).

The dominant heat sources at the 4.5~K stage are the FTS scanning mirror mechanism, achieved through a stiffness-compensated reaction-less design;\cite{Cournoyer2023} the two BSMs, building on cryogenic wheel-mechanism heritage from ISO;\cite{Bollinger1999} the ADR heat sink during re-magnetisation; the SVM harnesses; and parasitic radiative and conductive loads from the 25~K shield. The total stage load remains comfortably within the 4~K cooler capacity,\cite{ESA_M8F3_Call2025} with the detailed budget to be reported in a forthcoming paper.

\section{SUB-KELVIN COOLING CHAIN: MULTI-STAGE ADR}
\label{sec:subK}

Cooling below 4.5~K is provided by a multi-stage ADR derived from the design developed for the NewAthena/X-IFU instrument,\cite{Duval2024_5StageADR} also benefiting from SPICA/SAFARI\cite{Duval2015_SPICA_SAFARI_ADR} and LiteBIRD\cite{2020JLTP..199..730D} developments. It provides three successive stages: a \textbf{T2 continuous stage} at 1.8~K, which intercepts the dominant parasitic loads from the FPA support and maintains the BBIR within its 2.5--2.9~K range; a \textbf{T1 continuous stage} at 350~mK for intermediate interception; and a \textbf{T0 single-shot stage} reaching 50~mK, reaching 50~mK of continuous science operations during 28 with a duty cycle exceeding 80\%. This stage directly cools the detector units via a copper thermal strap. The demonstrated thermal stability of the X-IFU ADR\cite{Maisonnave2022_ADR} provides ample margin for stable KID operation at 50~mK.

\section{FOCAL PLANE ASSEMBLY THERMAL DESIGN}
\label{sec:fpa}

The FPA, whose multi-stage thermal isolation draws on the philosophy demonstrated on Planck HFI\cite{Planck_thermal2011} and recent structural and thermal model developments,\cite{Sauvage2025_StructuralThermalCCDR} houses four detector units (two Low-Frequency and two High-Frequency Detector Units), each consisting of a smoothwall multimoded feedhorn coupled to a segmented KID array. It is built around an isostatic hexapod surrounded by a 1.8~K radiation shield.

Maintaining all four detector units at 50~mK from the 4.5~K bench requires very effective thermal isolation through a multi-stage interception scheme. The primary structural supports between the 4.5~K bench and the 1.8~K interface are CFRP T700 struts, whose low thermal conductivity minimises the conductive load reaching the cold stages. The struts are split at the 1.8~K and 320~mK stages, with thermal collars and braided straps at each intercept point, preventing continuous conductive heat propagation from 4.5~K to 50~mK. Below 320~mK, two complementary phenomena provide additional passive isolation: all structural interfaces use Al-6061-T6, which becomes superconducting below 1.1~K and whose thermal conductivity then drops dramatically; and at the 320~mK--50~mK boundary, the Kapitza boundary resistance further reduces parasitic loads on the coldest stage. Copper thermal straps redirect heat from the structural supports to the ADR cold stages. The detailed conductive load budget, derived from dedicated cryogenic measurements, will be presented in a forthcoming paper.

\section{BLACKBODY INTERNAL REFERENCE: THERMAL CONSTRAINTS}
\label{sec:bbir}

In science observation modes, one FTS input arm is directed at the sky while the other is directed at the BBIR, which must represent a known, stable, nearly perfect blackbody emission at temperatures bracketing $T_0 \approx 2.7255$~K, with setpoints spanning 2.5--2.9~K. The assembly comprises a primary absorbing region actively controlled over 2.5--2.9~K and a secondary high-frequency calibration cavity, independently heated to $\sim$20~K during dedicated sequences; its detailed electromagnetic and mechanical design is beyond the scope of this preliminary overview. The driving thermal requirements are a temperature stability of a few $\mu$K over the FTS scan duration ($\sim$2~s), and a limited parasitic load to the 1.8~K ADR stage. The BBIR is thermally isolated from the 4.5~K bench via insulating struts, with an Al-6061-T6 substrate to minimise gradients. The absorbing surface design draws on heritage from ARCADE\cite{Fixsen2011} and the MetOp-SG MWI calibration target.\cite{Simonetto2021} Temperature is monitored by Cernox thermometers, with an absolute thermometry uncertainty of $\Delta T_0 \approx 100~\mu$K from ground calibration, sufficient for FOSSIL's measurement goals.

\section{THERMAL MARGINS}
\label{sec:budget}

A preliminary steady-state thermal model has been developed for the full chain, from the SVM panel at 293~K down to the 50~mK detector stage, accounting for the dominant radiative and conductive loads across each interception stage. At this early design stage the model retains only first-order contributions and conservative assumptions; it is intended to verify overall feasibility rather than to provide consolidated figures. The passive chain provides comfortable margins at the 130~K, 90~K and 50~K V-grooves, which can be further tuned through geometric adjustments of the V-groove surface area and inclination angles. On the active side, the 25~K shield, 4.5~K bench, and the 1.8~K, 320~mK and 50~mK ADR stages all close with positive margins against their respective cooling capacities. The complete quantitative budget, including the detailed load decomposition at each stage and the associated margins, will be presented in a forthcoming paper. The key outcome at this preliminary stage of the design is that the FOSSIL thermal architecture closes consistently, with positive margins at every temperature level, confirming its compatibility with the mission requirements.

\section{CONCLUSION}
\label{sec:conclusion}

The preliminary thermal architecture of FOSSIL presented here addresses an extraordinary engineering challenge: maintaining a complex optical instrument at 4.5~K and its detectors at 50~mK, while preserving a blackbody calibration reference stable to a few $\mu$K at $\sim$2.7~K, all within the resource constraints of an ESA medium-class mission. The baseline design outlined here demonstrates that this challenge is addressable through a staged passive cooling chain (MLI + 3 V-grooves + 25~K active shield) drawing on validated Planck and ARIEL heritage; an ESA-provided 4~K cryocooler bridging the SVM temperature to 4.5~K; a multi-stage ADR developed for NewAthena/X-IFU providing 1.8~K, 350~mK and 50~mK cooling; and a CFRP hexapod FPA support exploiting the superconducting transition of Al-6061-T6 and the low conductivity of CFRP to minimise thermal conductance at the coldest stages. The preliminary thermal model confirms positive margins at every temperature stage. While the present architecture remains a baseline concept to be consolidated and optimised during Phase~A, it benefits from a clear TRL development path: the multi-stage ADR will reach TRL-6 through NewAthena/X-IFU, and future instruments (BISOU,\cite{bisou2024} TMS,\cite{TMS2020} COSMO\cite{cosmo2024}) will further validate key sub-systems ahead of FOSSIL Phase~A, targeting a launch in the early 2040s. A dedicated paper presenting the full quantitative thermal budget and its experimental validation is in preparation.

\acknowledgments
The authors thank the full FOSSIL consortium (265 members in 15 countries) for their contributions to the mission concept.

\bibliographystyle{spiebib}
\bibliography{report}

\end{document}